\documentclass[twocolumn,prl,showpacs,superscriptaddress]{revtex4}

\usepackage{graphicx}
\usepackage{dcolumn}
\usepackage{bm}

\begin{document}

\title{The Second Landau Level Projection of Antiholomorphic Pfaffian Wavefunctions}
\author{Jian Yang}
\email{jyangmay1@yahoo.com}
\altaffiliation{Permanent address: 4610 Ravensthorpe Ct, Sugar Land, TX 77479, USA}

\begin{abstract}

We propose a new state described by the second Landau level (SLL) projection of a generalized Moore-Read Pfaffian wavefunction with an antiholomorphic pairing component. Unlike the PH-Pfaffian state which is described by the lowest Landau level (LLL) projection of the same antiholomorphic Pfaffian wavefunction, we find the proposed state, which we call the SLL PH-Pfaffian, represents a gapped, incompressible phase, and provides an excellent description for the exact ground state of finite-size SLL Coulomb-interacting electrons over a range of the short distance interaction strength. We also find the SLL PH-Pfaffian, when mapped to the LLL, has a very large overlap and the same low-energy orbital entanglement spectrum structures with the anti-Pfaffian state. We discuss possible interpretations of our results on the nature of the topological order of both states.

\end{abstract}
 \maketitle

The nature of the $\nu=5/2$ fractional quantum Hall effect (FQHE) \cite{Willett} has been under active study both theoretically and experimentally. The leading candidates for the ground state in numerical studies are the Pfaffian state\cite{MR} and its particle-hole (PH) conjugate, the anti-Pfaffian state\cite{Levin} \cite{Lee}. Since the Coulomb interaction projected into a single Landau level is PH symmetric, the Pfaffian and the anti-Pfaffian states are degenerate. The presence of the Landau level mixing the degeneracy is lifted and the anti-Pfaffian state is favored energetically\cite{EHRezayi}. On the other hand, another related state, termed PH-Pfaffian state\cite{Son}\cite{Zucker}, has recently attracted considerable attention. Although it has been shown to have high level of the PH symmetry\cite{Yang}\cite{Balram}\cite{Mishmash}, in no numerical work has it ever appeared as a ground state of a disorder-free system. 

Since the nature of these states is reflected in the gapless edge modes due to the "bulk-edge" correspondence, most experiments probe the edge physics. The Pfaffian state has two edge modes, a charged chiral boson mode and a neutral chiral Majorana fermion mode, both propagating in the downstream direction. The anti-Pfaffian state is a PH conjugate of the Pfaffian. According to the two composite edge effective field theory\cite{Levin} \cite{Lee}, an edge between an anti-Pfaffian bulk and vacuum can be formed by adding a narrow region of $\nu = 1$ separating the anti-Pfaffian bulk and the $\nu = 0$ vacuum. The edge between anti-Pfaffian bulk and vacuum can be decomposed into the edge between the anti-Pfaffian bulk and $\nu = 1$ Hall liquid, which is equivalently via PH symmetry to the edge between a Pfaffian bulk and $\nu = 0$ vacuum, and the edge between the $\nu = 1$ Hall liquid and vacuum. Therefore, the anti-Pfaffian state has a downstream charged boson edge mode, an upstream neutral Majorana edge mode, and an upstream neutral boson edge mode. The upstream neutral boson mode can further be fermionized into two upstream Majorana neutral edge modes\cite{Lee}, and we end up with three upstream Majorana neutral edge modes for the anti-Pfaffian state. On the other hand, the PH-Pfaffian state has two edge modes, a downstream charged boson mode and a upstream neutral Majorana fermion mode. Since a downstream boson edge mode contributes one unit ($\frac{\pi^2k^2_B}{3h}T$) to the thermal Hall conductance $\kappa_{xy}$, and a downstream (upstream) Majorana fermion mode contributes positive (negative) one half unit to $\kappa_{xy}$, we have $\kappa_{xy} = 1+1/2 = 3/2$ for the Pfaffian state, $\kappa_{xy} = 1-3/2 = -1/2$ for the anti-Pfaffian state,  and $\kappa_{xy} = 1-1/2 = 1/2$ for the PH-Pfaffian state. Experimentally\cite{Banerjee}, the thermal Hall conductance is found to be $\kappa_{xy} = 5/2$. Considering the filled lowest Landau level (LLL) of both spins contributes $2$ units to the thermal conductance, the experiment is only consistent with the PH-Pfaffian and appears to rule out both the Pfaffian and anti-Pfaffian. One explanation is the disorder induced formation of Pfaffian and anti-Pfaffian domains\cite{Mross}\cite{CWang}\cite{Lian}, although it is unclear whether this mechanism can account for the experimental observation\cite{Simon}\cite{Zhu}. Another possibility is incomplete thermal equilibration of an anti-Pfaffian edge\cite{Simon1}\cite{Asasi}. However, a more recent experiment\cite{Dutta}, probing the counterpropagating chirality of the Majorana mode by measuring partition noise, has further strengthened the case for the PH-Pfaffian and weakened its competitors, the Pfaffian and anti-Pfaffian.

In this paper we offer a possible resolution to the discrepancy between the experimental and the numerical results. We begin with the Moore-Read Pfaffian wavefunction\cite{MR}
\begin{equation}
\label{Pfaffian} {\Psi}_{Pf} = {Pf} ( \frac{1}{z_i-z_j } )   \prod\limits_{i<j}^N (z_i-z_j)^2
\end{equation}
where $z_j = x_j+iy_j$ is the complex coordinate of the $j_{th}$ electron, $N$ is the total number of electrons, and $Pf[A]$ is the Pfaffian of an antisymmetric matrix $A$. It is the holomorphic Pfaffian term that results in a downstream neutral Majorana fermion edge mode. 

In order to reverse the chirality of the neutral Majorana fermion mode as required by the PH-Pfaffian state, one can simply change the pairing component of Eq.(\ref{Pfaffian}) from holomorphic to antiholomorphic, and project it to the lowest Landau level (LLL)\cite{Zucker}:
\begin{equation}
\label{PH-Pfaffian} {\Psi}_{PH-Pf} = P_{LLL}{Pf} ( \frac{1}{z_i^*-z_j^* } )   \prod\limits_{i<j}^N (z_i-z_j)^2
\end{equation}
where $P_{LLL}$ is the LLL projection operator. Another form of the PH-Pfaffian wavefunction was also proposed\cite{Yang} (a related wavefunction was previously introduced in Ref.\cite{Jolicoeur}) by multiplying a stabilization factor\cite{Mishmash} $\prod\limits_{i<j}^N |z_i-z_j|^2$ before projecting it to the LLL:
\begin{equation}
\label{PH-Pfaffian1} {\Psi}_{PH-Pf}' = P_{LLL}{\Psi}_{Pf,b}^{\nu=-1}  \prod\limits_{i<j}^N (z_i-z_j)^3
\end{equation}
where ${\Psi}_{Pf,b}^{\nu=-1}$ is the $\nu=-1$ Bosonic Pfaffian wavefunction
\begin{equation}
\label{bPf} {\Psi}_{Pf,b}^{\nu=-1} = {Pf} ( \frac{1}{z_i^*-z_j^* } )   \prod\limits_{i<j}^N (z_i^*-z_j^*) 
\end{equation}
The LLL projection in Eq.(\ref{PH-Pfaffian1}) can be carried out by turning $z_i^*$ into derivative operator $\partial/{\partial z_i}$\cite{Yang}\cite{Jolicoeur}.

Now let's examine the underlying reasons for the LLL projection. In the presence of a strong magnetic field, it is a good approximation to restrict electrons in the LLL for the FQHE states at $\nu < 1$ filling factors. For $\nu = 1/m$ FQHE states with $m$ being odd integers, the Laughlin wavefunction\cite{Laughlin} already resides entirely in the LLL, no projection is required. For other $\nu < 1$ hierarchical FQHE states, the construction of the composite fermion wavefunctions\cite{Jain} necessarily involve high Landau levels, and therefore require an explicit LLL projection. On the other hand, for the FQHE involving the SLL such as $\nu = 5/2$ there is no underlying physical reason to use the LLL projection. To the contrary, for the $\nu = 5/2$ FQHE state, it is considered to be a good approximation to focus only on the half-filled SLL electron system, assuming the filled LLL of both spins to be inert. Nevertheless, it is in generally more convenient to work in the LLL first because the wavefunctions can normally be constructed in a much simpler form in the LLL, such as the Pfaffian wavefunction, and the corresponding SLL wavefunction can then be obtained by applying raising operators to the LLL wavefunction. 

None of the above reasons applies to the LLL projection of the PH-Pfaffian wavefunction Eq.(\ref{PH-Pfaffian}) or Eq.(\ref{PH-Pfaffian1}), as it is not required by the underlying physics, nor does it make the resulting wavefunction any simpler. The reason for the LLL projection is the PH symmetry requirement. On a sphere, the PH-Pfaffian wavefunction occurs at flux $N_{\phi}= 2N-1$, and the degeneracy of the LLL is $N_{\phi}+1 = 2N$, twice the number of electrons, which the necessary condition for the PH symmetry. However, this also turns out to be the very reason that the PH-Pfaffian wavefunction fails to represent a gapped, incompressible ground state. 

To that end it seems more naturally to propose the following wave function:
\begin{equation}
\label{SLL-PH-Pfaffian} {\Psi}_{SPH-Pf} = P_{SLL}{Pf} ( \frac{1}{z_i^*-z_j^* } )   \prod\limits_{i<j}^N (z_i-z_j)^2
\end{equation}
where $P_{SLL}$ is the SLL projection operator. We call the state the second Landau level PH-Pfaffian (SPH-Pfaffian) to distinguish it from the PH-Pfaffian state. The SPH-Pfaffian state is obviously not PH symmetric because the degeneracy of the SLL at $N_{\phi}= 2N-1$ is $N_{\phi}+3 = 2N+2$, as opposed to the degeneracy of the LLL which is $N_{\phi}+1 = 2N$. However, as we will show shortly this state does represents a gapped, incompressible phase. If we assume the SLL projection does not fundamentally alter the pairing structure of the original unprojected wavefunction, we expect the SPH-Pfaffian state also possesses the PH-Pfaffian topological order, and therefore consistent with the experiments. As in the case of the PH-Pfaffian state, we can have another form of the SPH-Pfaffian wavefunction:
\begin{equation}
\label{SLL-PH-Pfaffian1} {\Psi}_{SPH-Pf}' = P_{SLL}{\Psi}_{Pf,b}^{\nu=-1}  \prod\limits_{i<j}^N (z_i-z_j)^3
\end{equation}
by multiplying a stabilization factor $\prod\limits_{i<j}^N |z_i-z_j|^2$ before projecting it to the SLL.

In Haldane's spherical geometry\cite{Haldane}, Eq.(\ref{SLL-PH-Pfaffian}) can be written as:
\begin{equation}
\label{SSPH-Pfaffian} {\Psi}_{SPH-Pf} = P_{SLL}{Pf}(\frac{1}{  u_i^*v_j^*-u_j^*v_i^*})  \prod\limits_{i<j}^N (u_iv_j-u_jv_i)^{2} 
\end{equation}
where $(u, v)$ are the spinor variables describing electron coordinates. The $P_{SLL}$ can be carried by using the following equation\cite{Yutushui}:
\begin{equation}
\label{SLL1} \frac{1}{  u_i^*v_j^*-u_j^*v_i^*} = \sum_{l,m} (-1)^{m-\frac{1}{2}}\frac{8\pi}{2l+1}Y_{l,m}^{\frac{1}{2}}(u_i,v_i)Y_{l,-m}^{\frac{1}{2}}(u_j,v_j)
\end{equation}
where $Y_{l,m}^q$ is the monopole harmonics wavefunction\cite{Wu1}\cite{Wu2}, and equation\cite{Wu2} 
\begin{eqnarray}
\label{SLL2}
&& Y_{l,m}^{q}Y_{l',m'}^{q'} = \sum_{l"}(-1)^{l+l'+l"+2(q"+m")}(\frac{(2l+1)(2l'+1)}{4\pi(2l"+1)})^{\frac{1}{2}}
\nonumber \\
&&<lm,l'm'|l"m"><lq,l'q'|l"q">Y_{l",m"}^{q"}
\end{eqnarray}
where $q" = q+q'$ and $m" = m+m'$.

Similarly, on a sphere Eq.(\ref{SLL-PH-Pfaffian1}) can be written as:
\begin{equation}
\label{SSPH-Pfaffian1} {\Psi}_{SPH-Pf}' = P_{SLL}{\Psi}_{Pf,b}^{\nu=-1} \prod\limits_{i<j}^N (u_iv_j-u_jv_i)^{3} 
\end{equation}
where
\begin{equation}
\label{bPf} {\Psi}_{Pf,b}^{\nu=-1} = {Pf}(\frac{1}{  u_i^*v_j^*-u_j^*v_i^*}) \prod\limits_{i<j}^N (u_i^*v_j^*-u_j^*v_i^*)
\end{equation}
Since ${\Psi}_{SPH-Pf}'$ in Eq.(\ref{SSPH-Pfaffian1}) is the product of the $\nu=-1$ Bosonic Pfaffian wavefunction and the Laughlin wavefunction, we can expand the $\nu=-1$ Bosonic Pfaffian wavefunction in terms of $(Y_{l,m}^{\frac{N}{2}-1})^*$, and $\prod\limits_{i<j}^N (u_iv_j-u_jv_i)^{3}$ in terms of $Y_{l,m}^{(3N-3)/2}$, and finally using Eq.(\ref{SLL2}) and the following identity
\begin{equation}
\label{SLL3} (Y_{l,m}^{q})^* = (-1)^{q+m}Y_{l,-m}^{-q}
\end{equation}
to carry out the SSL projection $P_{SLL}$.

There seems to be no connection between the SPH-Pfaffian and the anti-Pfaffian state as one occurs at flux $N_{\phi}= 2N-1$ and the other occurs at flux $N_{\phi}= 2N+1$. However, since the SPH-Pfaffian is a SLL state and the anti-Pfaffian is a LLL state, we can map the SPH-Pfaffian to a LLL state with flux $N_{\phi}= 2N+1$ or map the anti-Pfaffian state to a SLL state with flux $N_{\phi}= 2N-1$ by identifying the single electron states of the LLL and those of the SLL with the same angular momentum. With this mapping, we can calculate the overlaps among the SPH-Pfaffian state, the anti-Pfaffian state, and the exact ground state of finite-size SLL Coulomb-interacting electrons. To be consistent with the previous numerical studies, we will use the LLL notation to present our numerical results. Specifically, the SPH-Pfaffian state refers to the LLL mapped SPH-Pfaffian state with flux $N_{\phi}= 2N+1$.

\begin{figure}[tbhp]
\label{fig:Overlap}
\includegraphics[width=\columnwidth]{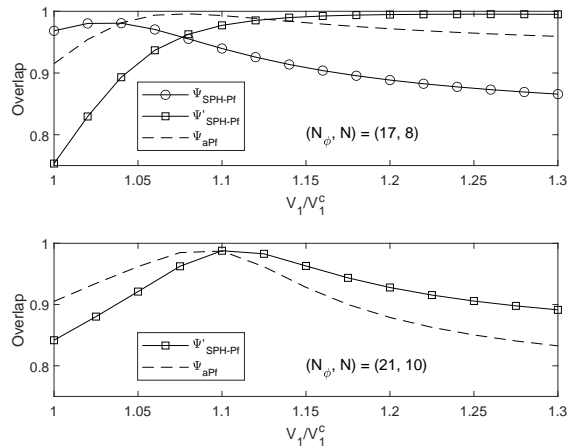}
\caption{\label{fig:Overlap} Overlaps of the SPH-Pfaffian state ${\Psi}_{SPH-Pf}$,  ${\Psi}_{SPH-Pf}'$, and the anti-Pfaffian state ${\Psi}_{aPf}$ respectively with the finite-size SLL Coulomb-interacting electrons as the function of the first Haldane pseudopotential $V_1$ normalized by its SLL Coulomb value $V_1^c$. The top panel is for a finite system $(N_{\phi}, N) = (17, 8)$, and the bottom panel for $(N_{\phi}, N) = (21, 10)$.}
\end{figure}

In Fig.~\ref{fig:Overlap}, we plotted the overlaps of the SPH-Pfaffian state ${\Psi}_{SPH-Pf}$ and ${\Psi}_{SPH-Pf}'$ described by Eq.(\ref{SSPH-Pfaffian}) and Eq.(\ref{SSPH-Pfaffian1}) respectively with the finite-size SLL Coulomb-interacting electrons in the spherical geometry. For comparison, we also plot the overlap between the anti-Pfaffian state ${\Psi}_{aPf}$ and the exact ground state. The anti-Pfaffian state ${\Psi}_{aPf}$ is obtained directly from the PH conjugate of the Pfaffian state. The top panel is for a finite system $(N_{\phi}, N) = (17, 8)$, and the bottom panel for $(N_{\phi}, N) = (21, 10)$. Please note we do not have the result of ${\Psi}_{SPH-Pf}$ for $(N_{\phi}, N) = (21, 10)$ as it becomes numerically intractable. The exact ground state is obtained using the SLL Coulomb interaction with the first Haldane pseudopotential ratios of $V_1/V_1^c$ ranging from $1$ to $1.3$, where $V_1^c$ is the SLL Coulomb value of $V_1$. We see over a certain range of $V_1/V_1^c$ around $1.1$, both forms of LLL mapped SPH-Pfaffian state and the anti-Pfaffian have large overlaps with the exact ground state. Compared to the anti-Pfaffian ${\Psi}_{aPf}$, the SPH-Pfaffian ${\Psi}_{SPH-Pf}$ favors smaller $V_1/V_1^c$, while ${\Psi}_{SPH-Pf}'$ favors larger $V_1/V_1^c$ due to the stabilization factor $\prod\limits_{i<j}^N |z_i-z_j|^2$. 

As seen from Fig.~\ref{fig:Overlap}, both forms of SPH-Pfaffian state ${\Psi}_{SPH-Pf}$ and ${\Psi}_{SPH-Pf}'$, and the anti-Pfaffian ${\Psi}_{aPf}$ have large overlaps with the exact ground state, it is not surprising that they have large overlaps among themselves as well (see Table I). 

\begin{table}[h]
\centering
\begin{tabular}{|c|c|c|c|}
\hline
$(N_{\phi}, N)$&$(13, 6)$ & $(17, 8)$ & $(21, 10)$ \\
\hline
  $<{\Psi}_{aPf}|{\Psi}_{SPH-Pf}>$&$0.9715$ & $0.9635$ & $$\\
\hline
  $<{\Psi}_{aPf}|{\Psi}_{SPH-Pf}'>$&$0.9842$ & $0.9538$ & $0.9786$\\
\hline
  $<{\Psi}_{SPH-Pf}|{\Psi}_{SPH-Pf}'>$&0.9184$$ & $0.8669$ & $$\\
\hline
\end{tabular}
\caption{\label{tab:table1}Overlaps among both forms of SPH-Pfaffian state ${\Psi}_{SPH-Pf}$ and ${\Psi}_{SPH-Pf}'$, and the anti-Pfaffian ${\Psi}_{aPf}$ for 3 finite systems $(N_{\phi}, N) = (13, 6), (17, 8)$, and $(21, 10)$. For $(N_{\phi}, N) = (21, 10)$, no result available for ${\Psi}_{SPH-Pf}$ as it becomes numerically intractable.}
\end{table}

In view of such large overlaps between the anti-Pfaffian state and the SPH-Pfaffian state, we expect their low-energy level entanglement spectra\cite{Li} should match well. This is indeed the case. In Fig.~\ref{fig:Entanglement}, we perform an orbital decomposition where subsystem $A$ contains half of the electrons with positive angular momentum $Lz$ and $B$ the other half with negative $Lz$, and plot the corresponding orbital entanglement spectrum as a function of the total angular momentum in subsystem $A$ for a finite system ($N_{\phi}, N$) = ($21, 10$). The top panel is for the the SPH-Pfaffian state ${\Psi}_{SPH-Pf}'$, and the bottom panel is for the anti-Pfaffian state ${\Psi}_{aPf}$. The multiplicities of the $4$ lowest entanglement levels at $L_z^A = 24.5, 25.5, 26.5, 27.5$ are $1, 1, 3$, and $5$, which are the same for ${\Psi}_{SPH-Pf}$ and ${\Psi}_{aPf}$.
\begin{figure}[tbhp]
\label{fig:Entanglement}
\includegraphics[width=\columnwidth]{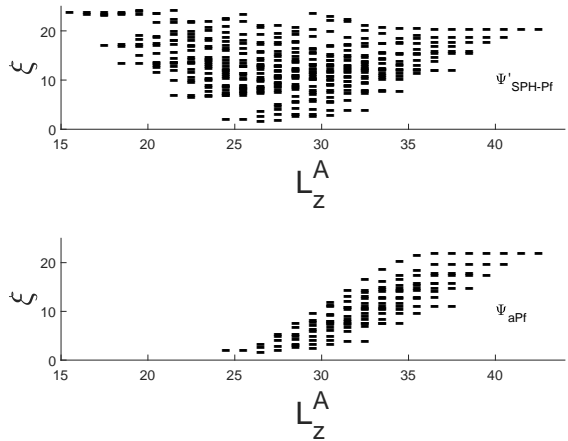}
\caption{\label{fig:Entanglement} Orbital entanglement spectrum as a function of the total angular momentum $Lz$ in subsystem $A$ for a finite system  ($N_{\phi}, N$) = ($21, 10$). The top panel is for the SPH-Pfaffian state ${\Psi}_{SPH-Pf}'$, and the bottom panel is for the anti-Pfaffian state ${\Psi}_{aPf}$. }
\end{figure}

Now the question is how can the two states, the SPH-Pfaffian and the anti-Pfaffian, which apparently possess different topological orders, have such a remarkable large overlap and the same or very similar low-energy orbital entanglement spectrum structure? The answer to the question is obviously critical to explain the experimental results. One possibility is that the SLL projection $P_{SLL}$ fundamentally alters the pairing structure, changing the SPH-Pfaffian state from having a single upstream neutral Majorana edge mode to three upstream neutral Majorana edge mode, making the SPH-Pfaffian state the same topological order as the anti-Pfaffian state. Although we cannot rule out this possibility, it seems more plausible the other way around. One scenario is that the upstream neutral boson edge mode of the ant-Pfaffian state according to the two composite edge effective field theory\cite{Levin} \cite{Lee}, is in fact a gapped mode, and will not be seen in the experiment. This will make the anti-Pfaffian state effectively exhibit the same topological order as the SPH-Pfaffian from the experimental perspective. Another possibility is that the large overlap and the same low-energy entanglement spectrum structure are just a finite size effect. In the thermodynamic limit, the SPH-Pfaffian and the anti-Pfaffian represent different topological phases, and the SPH-Pfaffian might be better for some range of parameters close to that of the experiment. More works are required to validate or invalidate each of the possibilities.  

In summary, the difference between the newly proposed SLL-PH-Pfaffian and the previously studied PH-Pfaffian state lies in the different Landau level projections.  Unlike the PH-Pfaffian state, the SLL-PH-Pfaffian state is not PH symmetric but has a very large overlap with the exact ground state of finite-size systems of the SLL Coulomb-interacting electrons, and therefore does represent a gapped, incompressible phase. However, its large overlap and the same low-energy orbital entanglement spectrum structure with the anti-Pfaffian state poses an open question as to the nature of the topological order of both states.

\end{document}